\documentclass[conference]{IEEEtran}
\IEEEoverridecommandlockouts
\usepackage{cite}
\usepackage{amsmath,amssymb,amsfonts}
\usepackage{algorithmic}
\usepackage{graphicx}
\usepackage{textcomp}
\usepackage{xcolor}
\def\BibTeX{{\rm B\kern-.05em{\sc i\kern-.025em b}\kern-.08em
    T\kern-.1667em\lower.7ex\hbox{E}\kern-.125emX}}

\usepackage{hyperref}
\usepackage[capitalise,noabbrev]{cleveref}  % nameinlink

\usepackage{stmaryrd}  % font with \shortdownarrow \shortuparrow
\usepackage{dsfont}

\usepackage{multirow}
\usepackage{tabularx,booktabs} %for professional tables
\usepackage{makecell}  % for newlines in table cells
\usepackage{pifont}  % for cmark and xmark
\newcommand{\cmark}{\ding{51}}%
\newcommand{\xmark}{\ding{55}}%

\usepackage{subfigure}

\usepackage{tcolorbox}  % for color boxes

\usepackage{caption}
\usepackage{tikz}
\usetikzlibrary{arrows.meta}
\usetikzlibrary{decorations.pathreplacing,calligraphy}

\newcommand{\darkGreen}{green!80!red!70!black}
\newcommand{\violet}{blue!70!red!99!black}
\newcommand{\bone}{green!35!blue!20!white!99!black}

\newcommand{\paramColor}{\bone}
\newcommand{\nonParamColor}{white}

\newcommand{\boxFont}{\ttfamily}  % sffamily rmfamily ttfamily
\newcommand{\arrowFont}{\boxFont \small}

\newcommand{\F}{\textcolor{\darkGreen}{\textbf{F}}}
\newcommand{\K}{\textcolor{\violet}{\textbf{K}}}
\newcommand{\FW}{\F$\times$T'}
\newcommand{\KW}{\K$\times$T'}
\newcommand{\oneW}{\textbf{1}$\times$T'}

\tikzset{
  >={Latex[width=2mm,length=2mm]},
  base/.style = {rectangle, draw=black,
    fill=white, minimum width=2mm, minimum height=7mm,
    text centered, font=\boxFont},
  baseround/.style = {rectangle, draw=black, rounded corners,
    fill=white, minimum width=10mm, minimum height=0.7cm,
    text centered, font=\boxFont},
  basecircle/.style = {circle, draw=black,
    fill=white, minimum width=2mm,
    text centered, font=\boxFont},
  txt/.style = {base, draw=white, fill=white, minimum width=5mm, minimum height=5mm},
  param/.style = {baseround, fill=\paramColor},
  nonparam/.style = {baseround, fill=\nonParamColor},
}

\usepackage{acronym}
\acrodef{PPT}[PPT]{Polyphonic Piano Transcription}
\acrodef{MIREX}[MIREX]{Music Information Retrieval Evaluation eXchange}

\acrodef{DL}[DL]{Deep Learning}
\acrodef{OF}[\textsc{O\&F}]{\textsc{Onsets\&Frames}}
\acrodef{OV}[\textsc{O\&V}]{\textsc{Onsets\&Velocities}}
\acrodef{DNN}[DNN]{Deep Neural Network}
\acrodef{CNN}[CNN]{Convolutional Neural Network}
\acrodef{ResNet}[ResNet]{Residual Neural Network}
\acrodef{CAM}[CAM]{Context-Aware Module}
\acrodef{RNN}[RNN]{Recurrent Neural Network}
\acrodef{LSTM}[LSTM]{Long-Short Term Memory}
\acrodef{BCE}[BCE]{Binary Cross-Entropy}
\acrodef{BN}[BN]{Batch Normalization}
\acrodef{SBN}[SBN]{Sub-Spectral Batch Normalization}
\acrodef{STFT}[STFT]{Short-Term Fourier Transform}
\begin{document}

\title{Onsets and Velocities: Affordable Real-Time Piano Transcription Using Convolutional Neural Networks\\
%% {\footnotesize \textsuperscript{*}Note: Sub-titles are not captured in Xplore and should not be used}
  \thanks{
    Work done as independent researcher at the \textit{IAMúsica} project supported by the \textit{Institut d'Estudis Baleàrics}, Balearic Islands.}
}

\author{\IEEEauthorblockN{Andres Fernandez}
  \IEEEauthorblockA{\textit{University of Tübingen and IMPRS-IS} \\
%% \textit{name of organization (of Aff.)}\\
Tübingen, Germany \\
a.fernandez@uni-tuebingen.de}
%% \and
%% \IEEEauthorblockN{Given Name Surname}
%% \IEEEauthorblockA{\textit{dept. name of organization (of Aff.)} \\
%% \textit{name of organization (of Aff.)}\\
%% City, Country \\
%% email address or ORCID}
%% \and
%% \IEEEauthorblockN{3\textsuperscript{rd} Given Name Surname}
%% \IEEEauthorblockA{\textit{dept. name of organization (of Aff.)} \\
%% \textit{name of organization (of Aff.)}\\
%% City, Country \\
%% email address or ORCID}
%% \and
%% \IEEEauthorblockN{4\textsuperscript{th} Given Name Surname}
%% \IEEEauthorblockA{\textit{dept. name of organization (of Aff.)} \\
%% \textit{name of organization (of Aff.)}\\
%% City, Country \\
%% email address or ORCID}
%% \and
%% \IEEEauthorblockN{5\textsuperscript{th} Given Name Surname}
%% \IEEEauthorblockA{\textit{dept. name of organization (of Aff.)} \\
%% \textit{name of organization (of Aff.)}\\
%% City, Country \\
%% email address or ORCID}
%% \and
%% \IEEEauthorblockN{6\textsuperscript{th} Given Name Surname}
%% \IEEEauthorblockA{\textit{dept. name of organization (of Aff.)} \\
%% \textit{name of organization (of Aff.)}\\
%% City, Country \\
%% email address or ORCID}
}

\maketitle

\begin{abstract}
  \acl{PPT} has recently experienced substantial progress, driven by the use of sophisticated \acl{DL} approaches and the introduction of new subtasks such as note onset, offset, velocity and pedal detection. This progress was coupled with an increased complexity and size of the proposed models, typically relying on non-realtime components and high-resolution data. In this work we focus on \textit{onset} and \textit{velocity} detection, showing that a substantially smaller and simpler convolutional approach, using lower temporal resolution (24ms), is still competitive: our proposed \acf{OV} model achieves state-of-the-art performance on the MAESTRO dataset for onset detection (F\textsubscript{1}=96.78\%) and sets a good novel baseline for onset+velocity (F\textsubscript{1}=94.50\%), while having $\sim$3.1M parameters and maintaining real-time capabilities on modest commodity hardware. We provide open-source code to reproduce our results and a real-time demo with a pretrained model \footnote{Setup \url{https://github.com/andres-fr/iamusica_training} \\ \hspace*{11.5pt}Demo: \url{https://github.com/andres-fr/iamusica_demo}}.
\end{abstract}


%%%%%%%%%%%%%%%%%%%%%%%%%%%%%%%%%%%%%%%%%%%%%%%%%%%%%%%%%%%%%%%%%%%%%%%%%%%%%%%%%%%%%%%%%%%%%%%%%%%%%%
%%%%%%%%%%%%%%%%%%%%%%%%%%%%%%%%%%%%%%%%%%%%%%%%%%%%%%%%%%%%%%%%%%%%%%%%%%%%%%%%%%%%%%%%%%%%%%%%%%%%%%
\section{Introduction}
\acresetall  % reset all acronyms from abstract

\subsection{Polyphonic Piano Transcription}

The task of \acfi{PPT} is useful for downstream tasks like musical analysis and resynthesis. Consider an audio {\it waveform} $x(t) \in \mathbb{R}^T$ with time $T$ that corresponds to a piano performance of a {\it score} $\mathcal{S}$; then the task of \ac{PPT} is to recover $\mathcal{S}$ from $x$. Here, $\mathcal{S}$ is a collection of $N$ {\it note events} $\{\mathcal{N}_n := (k_n, v_n, \shortdownarrow_n, \shortuparrow_n)\}_{n=1}^N$, where $k \in \{1, \dots, K \}$ specifies the {\it key} (typically $K=88$). The value $v \in [0, 1]$ indicates the intensity of the event (also called key {\it velocity}). The key {\it onset} (pressing) and {\it offset} (releasing) timestamps are specified by  $\shortdownarrow$ and $\shortuparrow$, respectively, where $0 \leq \shortdownarrow_n < \shortuparrow_n \leq T~~\forall n$.

\vspace{3pt}
There has been extensive effort in automating \ac{PPT}, typically articulated through challenges like the popular \acfi{MIREX} \cite{ismir2009} and featuring different techniques like handcrafted features, spectrogram factorization, probabilistic models \cite{amt13,driedger} and, more recently, \acfi{DL} \cite{dl_nature}. \ac{PPT} is typically evaluated by comparing the recovered score $\hat{\mathcal{S}}$ with the ground truth $\mathcal{S}$ on a test set, in an event-wise manner. Prominent efforts in curating datasets like MAPS\cite{maps,a_maps}, SMD\cite{smd} and MusicNet\cite{musicnet} were affected by imprecise annotations, insufficient training data, unrealistic interpretations and/or constrained recording conditions, which made evaluation more difficult and impeded the establishment of a unified benchmark for \ac{PPT}\cite{maestro}. The introduction of the MAESTRO dataset \cite{maestro} addressed many of these issues, by providing $\sim$200 hours of precisely annotated, high-quality audio data, encompassing a large variety of virtuosistic compositions, pianists and recording conditions, and incorporating evaluation splits. As a result, it quickly became a popular benchmark. Still, all pianos in MAESTRO are fairly similar: to capture more general settings and satisfy the ever-growing demand for training data, \cite{giantmidi} curated the GiantMIDI dataset, by sourcing over 1000 hours of piano music from YouTube and annotating them using \ac{DL}\cite{kong21}.

\begin{figure}[t]
  \centerline{\includegraphics[width=0.99\columnwidth]{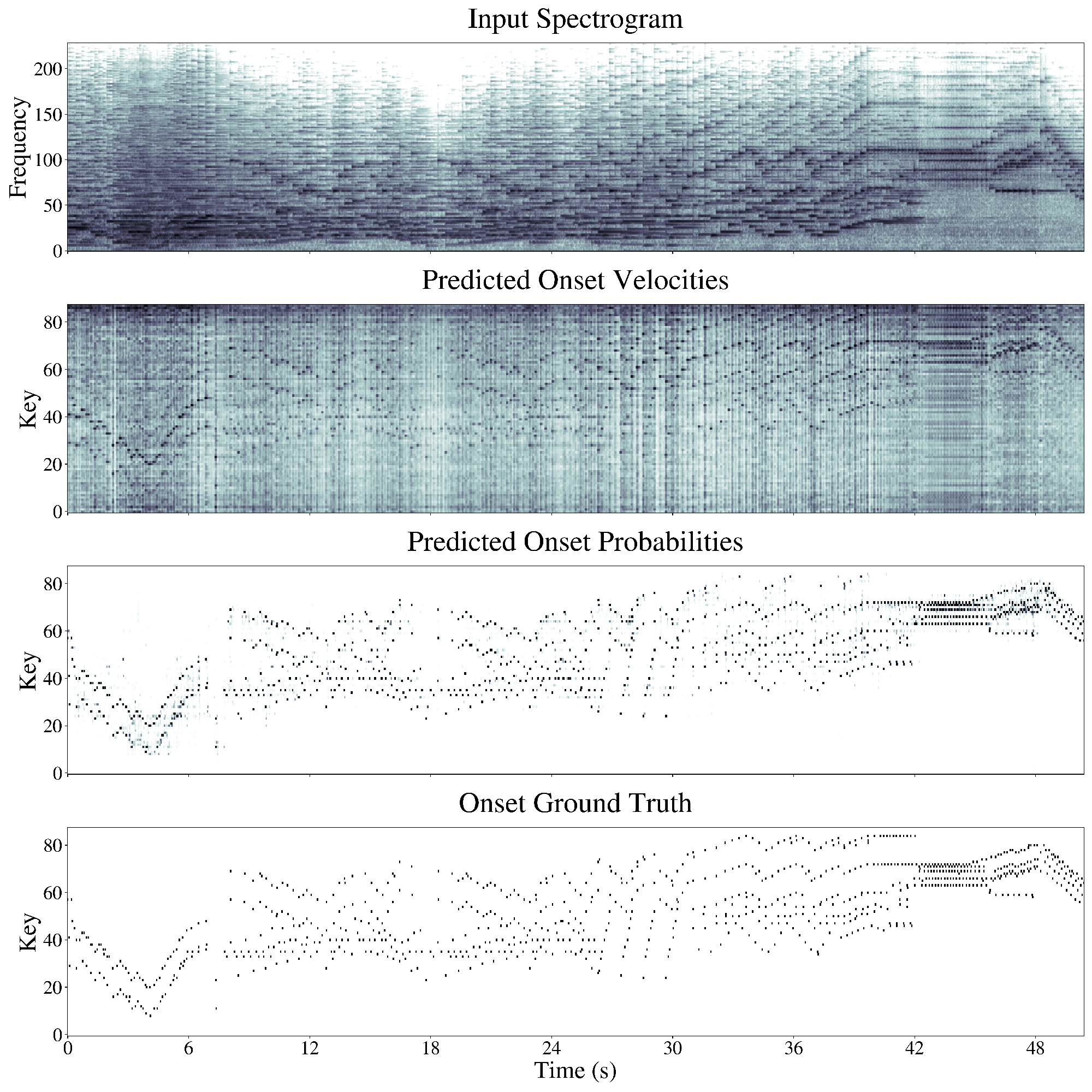}}
  \vspace{-2mm}
  \caption{
    Log-mel spctrogram ($X$) of a virtuosistic excerpt from the MAESTRO test set, followed by the corresponding velocity ($\hat{\mathcal{R}}_V$) and last-stage onset ($\hat{\mathcal{R}}_{\shortdownarrow}^{(3)}$) predictions, as well as the ground truth {\it piano roll} $\mathds{1}_{\shortdownarrow_3}$ (see \cref{sec:methodology} for details).\vspace{-4mm}
}
  \label{fig:qualitative}
\end{figure}

\subsection{The State of the Art in PPT}

One influential effort applying \ac{DL} to \ac{PPT} was \cite{cnn_maps16}. Their work cemented the following main trends:

\vspace{3pt}
\paragraph{Spectrograms} Despite promising efforts to use the $x(t)$ waveforms as \ac{DL} inputs \cite{leaf},  time-frequency representations like the spectrogram\cite[ch.19]{bracewell} remain competitive for \ac{PPT} and discriminative audio tasks in general \cite{panns,kong21}.
\paragraph{{\it Piano roll} supervision} Consider an alternative representation of $\mathcal{S}$, called {\it piano roll} $\mathcal{R} \in [0, 1]^{K \times T}$ (see \cref{fig:qualitative}), where entries $\mathcal{R}(k, t)$ encode the activity of channel $k$ at time $t$ (zero if inactive). This type of supervision consists in training the model to output a piano roll $\hat{\mathcal{R}}$ that predicts some ground truth $\mathcal{R}$, by minimizing the binary cross-entropy loss: $l_{BCE}(\mathcal{\mathcal{R}}, \hat{\mathcal{\mathcal{R}}}) \! = \! \langle \mathcal{\mathcal{R}}, -log(\hat{\mathcal{\mathcal{R}}}) \rangle \! + \!  \langle (1 \! - \! \mathcal{\mathcal{R}}), -log(1 \! - \! \hat{\mathcal{\mathcal{R}}}) \rangle$. Often, the ground truth is binarized and we have $\mathds{1} \in \{ 0, 1 \}^{K \times T}$ instead of $\mathcal{R}$. This approach requires to {\it decode} the predicted piano roll $\hat{\mathcal{R}}$  to obtain the event-based representation $\hat{\mathcal{S}} = dec_{H}(\hat{\mathcal{R}})$, typically by using a heuristic $H$, e.g. grouping consecutive active frames into singe notes.

\paragraph{Computer Vision} \ac{PPT} can be tackled effectively by treating spectrograms and piano rolls as images, and models like \acp{CNN} work well with minor adaptions.  % \cite{cnn98}

\vspace{3pt}
A major turning point was \ac{OF} \cite{onsets_frames}, which uses a sub-network to first predict a piano roll $\hat{\mathcal{R}}_{\shortdownarrow}$ encoding the probability of an {\it onset} (trained with a mask $\mathds{1}_{\shortdownarrow}$ that is active only when a key is pressed), and then uses another subnetwork to predict the {\it frames} $\hat{\mathcal{R}}_{\mathcal{N}}$ conditioned on $\hat{\mathcal{R}}_{\shortdownarrow}$ (trained with a mask $\mathds{1}_{\mathcal{N}}$, active for the whole duration of each note). \ac{OF} is then trained jointly via a multi-task loss $l_{BCE}(\mathds{1}_{\shortdownarrow}, \hat{\mathcal{R}}_{\shortdownarrow}) + l_{BCE}(\mathds{1}_{\mathcal{N}}, \hat{\mathcal{R}}_{\mathcal{N}})$. \ac{OF} achieved a steep improvement in all \ac{PPT} benchmarks, and also introduced a novel subtask, {\it note velocity}, modelled with a third sub-network that predicts a velocity piano roll $\hat{\mathcal{\mathcal{R}}}_{V}$ trained via masked $\ell_2$-norm loss $l_{V} = \langle \mathds{1}_{\shortdownarrow}, (\mathcal{\mathcal{R}}_{V} - \hat{\mathcal{\mathcal{R}}}_{V})^2 \rangle$. Due to its unprecedented effectiveness and versatility, \ac{OF} became a popular baseline \cite{adv_of, of_analysis}, but this was at the expense of increased complexity, including more elaborate decoder heuristics, a larger model, and the incorporation of bi-RNN layers\cite{lstm,birnn}, which preclude real-time applications.

\vspace{3pt}
More recently, \cite{kong21} pointed out issues with temporal precision on piano rolls, incorporating a trainable \textsc{Regression} model to enhance precision. They further expanded the model including sustain pedal detection capabilities. In \cite{hawthorne_transformer}, an {\it off-the-shelf} \textsc{Transformer}\cite{transformers} setup was used to produce $\hat{\mathcal{S}}$ directly from spectrograms in an end-to-end fashion. Apart from their good performance, both systems have in common their substantial size, increased time resolution and replacing decoder heuristics with an end-to-end differentiable solution, suggesting that decoding is a performance bottleneck.

\subsection{Proposed Contribution for PPT}

These state-of-the-art improvements in performance came entangled with increased complexity in the form of larger models, additional components and new sub-tasks\cite{of_analysis}. Understanding and disentangling this complexity is an active field of research: Alternative \ac{PPT} sub-task factorizations that do not rely on \ac{OF} were proposed, like nonlinear denoising vs. linear demixing \cite{kelz21}, sound source vs. note arrangement \cite{dds_overfitting} and ADSR envelopes \cite{adsr_kelz}. General approaches like using invertible neural networks\cite{invertible}, reconstruction tasks \cite{no_subtask} and additive attention \cite{of_analysis} were also explored.

\vspace{3pt}
In this work, we pursue the orthogonal goal of achieving real-time capabilities. For that, we observe that the masked loss $l_V$ imposes time-locality around the onsets, and follow up on several ideas: the importance of the onsets \cite{onsets_frames} as well as decoder heuristics\cite{of_analysis}, and the idea that note velocity is naturally associated with the onset \cite{kong21}. We propose that {\it  a convolutional end-to-end method for onsets and velocities leads to efficiency gains and affordable real-time capabilities without compromising performance}, and that {\it efficient decoding heuristics replace the need for high temporal resolution and complex inference schemes}.

%% \vspace{3pt}
%% \begin{tcolorbox}[
%%     colback=\bone!15!white, colframe=\bone!95!gray, coltitle=black,
%%     boxrule=0.5pt,
%%     left=0pt, right=0pt, top=0pt, bottom=0pt,
%%     title={We present \ac{OV}, featuring:}
%%   ] % sharp corners
%%   \begin{enumerate}
%%   \item State-of-the-art performance for {\it onset} detection and a good baseline for {\it onsets+velocities} on MAESTRO.
%%   \item A substantially smaller \ac{CNN} and a simple decoder, enabling affordable real-time inference on CPU.
%% \item A multi-task training scheme without any data augmentations or extensions, based on piano rolls at 24ms resolution.
%% \end{tcolorbox}
%% \end{enumerate}

\vspace{0.7em}
\textbf{We present \ac{OV}, featuring:}
\begin{enumerate}
\item State-of-the-art performance for {\it onset} detection and a good baseline for {\it onsets+velocities} on MAESTRO.
\item A substantially reduced \ac{CNN} (no recurrent layers) based on piano rolls at 24ms resolution, enabling affordable real-time inference on modest hardware.
\item A straightforward decoding mechanism, enabling a multi-task training scheme without any data augmentations or extensions.
\end{enumerate}
\vspace{0.3em}

In \cref{sec:methodology} we present our \ac{OV} method. \cref{sec:experiments} presents experiments substantiating our claims. We also provide a PyTorch \cite{pytorch} open-source implementation with a real-time demo. \cref{sec:conclusion} concludes and proposes future work.

%%%%%%%%%%%%%%%%%%%%%%%%%%%%%%%%%%%%%%%%%%%%%%%%%%%%%%%%%%%%%%%%%%%%%%%%%%%%%%%%%%%%%%%%%%%%%%%%%%%%%%
%%%%%%%%%%%%%%%%%%%%%%%%%%%%%%%%%%%%%%%%%%%%%%%%%%%%%%%%%%%%%%%%%%%%%%%%%%%%%%%%%%%%%%%%%%%%%%%%%%%%%%
\section{Methodology}\label{sec:methodology}

%%%%%%%%%%%%%%%%%%%%%%%%%%%%%%%%%%%%%%
\subsection{Model} \label{subsec:model}
%%%%%%%%%%%%%%%%%%%%%%%%%%%%%%%%%%%%%%
Given a waveform $x(t) \in \mathbb{R}^T$ at 16kHz, we compute its \ac{STFT}\cite{bracewell} with a Hann window of size 2048, and a hop size $\delta$=384 (i.e. a time resolution of $\Delta_t$=24ms). We then map it to 229 mel-frequency bins \cite{mel} in the 50Hz-8000Hz range, and take the logarithm, yielding our input representation: a {\it log-mel spectrogram} $X(f, t') \in \mathbb{R}^{229 \times T'}$, where $T' = \frac{T}{\delta}$ is the resulting ``compact'' time domain (see \cref{fig:qualitative}). We also compute the first time-derivative $\dot{X}(f, t') := X(f, t') - X(f, t' - 1)$ and concatenate it to $X$, forming the \ac{CNN} input. Using the same $\Delta_t$, we time-quantize the MIDI annotations into a piano roll $\mathcal{R}_{V} \in [0, 1]^{88 \times T'}$, where $\mathcal{R}_{V}(k_n, t_n')$ contains the velocity if key $k_n$ was pressed at time $\Delta_t t_n' \! \pm \! \frac{\Delta t}{2}$, and zero otherwise. We further binarize $\mathcal{R}_{V}$, yielding $\mathds{1}_{\shortdownarrow}$.

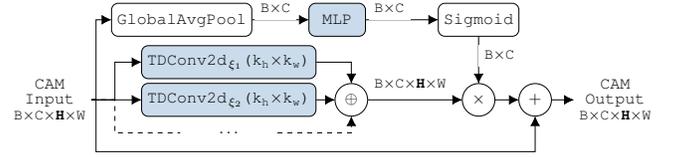
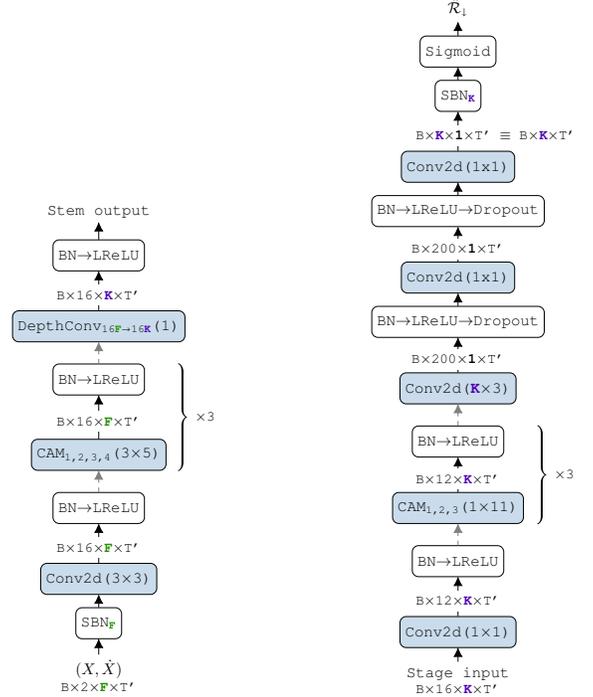
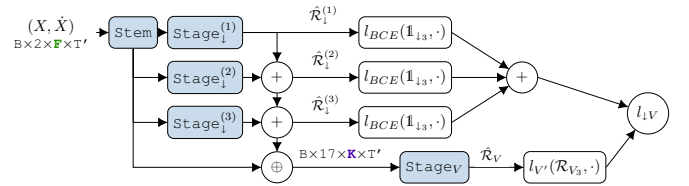
\begin{figure}[!htb]
  \centering
  \begin{tabular}[t]{c}
    \subfigure[
      Diagram of a \texttt{\textbf{CAM}}$_{\xi_1, \xi_2, \dots}(k_h \times k_w)$, based on \cite{cam}. The middle branch concatenates ($\oplus$) multiple time-dilated convolutions (\texttt{TDConv}), each with time dilation $\xi$. The output channels of each \texttt{TDConv} is \texttt{\textbf{H}} divided by the number of \texttt{TDConv} layers being concatenated, and padding is adjusted so shape is preserved. The top branch acts as a channel-wise attention mechanism ($\times$), featuring a Multi-Layer Perceptron (\texttt{MLP}) acting as a bottleneck (we use 2 layers with ReLU activation and 8 hidden dimensions). The bottom branch is a residual connection ($+$).\vspace{-4mm}]{
      \resizebox{1\linewidth}{!}{
        \begin{tikzpicture}[node distance=3cm,
  every node/.style={fill=white, font=\sffamily},
  align=center]
  \node (input)[txt]{CAM\\\\[1mm]Input\\\\[1mm]{\arrowFont B$\times$C$\times$\textbf{H}$\times$W}};
  \node (tdconv1)[param, right of=input, xshift=8mm] {TDConv2d\textsubscript{$\,\xi_2$}(k\textsubscript{h}$\times$k\textsubscript{w})};
  \node (tdconv2)[param, above of=tdconv1, yshift=-2.2cm] {TDConv2d\textsubscript{$\,\xi_1$}(k\textsubscript{h}$\times$k\textsubscript{w})};
  \node (tdconvdots)[txt, below of=tdconv1, xshift=0cm, yshift=23mm] {$\qquad\dots\qquad$};

  \node (pool)[nonparam, above of=input, xshift=28mm, yshift=-1.3cm] {GlobalAvgPool};
  \node (MLP)[param, right of=pool, xshift=3mm, yshift=0cm, minimum width=12mm] {MLP};
  \node (sigm)[nonparam, right of=MLP, xshift=0mm, yshift=0cm] {Sigmoid};
  \node (cat)[basecircle, right of=tdconv1, xshift=-4mm, yshift=0cm] {$\oplus$};
  \node (att)[basecircle, below of=sigm, xshift=0mm, yshift=1.3cm] {$\times$};
  \node (res)[basecircle, right of=att, xshift=-18mm, yshift=0cm] {$+$};
  \node (output)[txt, right of=res, xshift=-13mm, yshift=0cm] {CAM\\\\[1mm]Output\\\\[1mm]{\arrowFont B$\times$C$\times$\textbf{H}$\times$W}};

  \node (bottomMargin)[txt, below of=input, yshift=17mm]{};

  \newcommand{\innerDist}{5mm}
  \newcommand{\outerDist}{1mm}
  % residual arrow
  \draw [->] (input.east) -- ++(\outerDist, 0mm) --  ++(0mm,-11mm)  -| (res);

  % arrows left of conv
  \draw [->] (input.east) --++(0:\innerDist)|- (tdconv1);
  \draw [->] (input.east) --++(0:\innerDist)|- (tdconv2);
  \draw [dashed] (input.east) --++(0:\innerDist) |- (tdconvdots);
  % arrows right of conv
  \draw [->] (tdconv1.east) -- (cat);
  \draw [->] (tdconv2.east) -| (cat);
  \draw [->, dashed] (tdconvdots.east) -| (cat);
  % SE arrows
  \draw [->] (input.east) --++(0:\outerDist) |- (pool);
  \draw [->] (pool.east) |- node [above=0mm, xshift=5mm] {\arrowFont B$\times$C} (MLP);
  \draw [->] (MLP.east) |- node [above=0mm, xshift=5mm] {\arrowFont B$\times$C} (sigm);
  \draw [->] (sigm.south) -| node [right, yshift=-4mm] {\arrowFont B$\times$C} (att);
  % concat and attention
  \draw [->] (cat.east) -- node [above=1mm,xshift=-1mm,midway] {\arrowFont B$\times$C$\times$\textbf{H}$\times$W} (att);
  \draw [->] (att.east) -- (res);
  % output
  \draw [->] (res.east) -- (output);
\end{tikzpicture}
      }
      \label{fig:model1}
    }
    \\
    \begin{tabular}[t]{cc} % if you add [t], than sub images are pushed down
      \subfigure[Diagram of the \texttt{\textbf{Stem}}, which is a \ac{CAM}-powered residual \ac{CNN}. The \texttt{SBN} has \texttt{\F} frequency bands (one per vertical dimension). The $\times$\texttt{3} braces indicate 3 sequential blocks. Note the single-step \texttt{\F}$\shortrightarrow$\texttt{\K} transition via depthwise convolution \texttt{DepthConv(1)} with temporal kernel width of 1.]{
        \resizebox{0.45\linewidth}{!}{\newcommand{\horizMarginStem}{0mm}

\begin{tikzpicture}[node distance=3cm,
  every node/.style={fill=white, font=\sffamily},
  align=center]

  \node (input)[txt, minimum height=2mm]{$(X, \dot{X})$\\[1mm]{\arrowFont B$\times$2$\times$\FW}};

  \node (sbn)[nonparam, above of=input, yshift=-18mm]{SBN\textsubscript{\textcolor{\darkGreen}{\textbf{F}}}};

  \node (conv3x3)[param, above of=sbn, yshift=-20mm]{Conv2d(3$\times$3)};

  \node (bnrelu1)[nonparam, above of=conv3x3, yshift=-14mm]{BN$\rightarrow$LReLU};

  \node (cam1)[param, above of=bnrelu1, yshift=-18mm]{CAM\textsubscript{1,2,3,4}(3$\times$5)};
  \node (bnrelu2)[nonparam, above of=cam1, yshift=-13mm]{BN$\rightarrow$LReLU};

  \node (dconv)[param, above of=bnrelu2, yshift=-18mm]{DepthConv\textsubscript{16\F$\shortrightarrow$16\K}(1)};

  \node (bnrelu3)[nonparam, above of=dconv, yshift=-14mm]{BN$\rightarrow$LReLU};

  \node (output)[txt, above of=bnrelu3, yshift=-20mm,minimum height=2mm]{Stem output};

  \node (bottomMargin)[txt, below of=input, yshift=24mm]{};
  \node (leftMargin)[txt, left of=input, xshift=-\horizMarginStem]{};
  \node (rightMargin)[txt, right of=input, xshift=\horizMarginStem]{};

  % main arrow path
  \draw [->] (input.north) -- (sbn);

  \draw [->] (sbn.north) -- (conv3x3);
  \draw [->] (conv3x3.north) -- node[yshift=-1mm]{\arrowFont B$\times$16$\times$\FW} (bnrelu1);
  \draw [->,dashed,draw=black!50] (bnrelu1.north) -- (cam1);
  \draw [->] (cam1.north) -- node[yshift=-1mm]{\arrowFont B$\times$16$\times$\FW} (bnrelu2);

  \draw [->,dashed,draw=black!50] (bnrelu2.north) -- (dconv);

  \draw [->] (dconv.north) -- node[yshift=-1mm]{\arrowFont B$\times$16$\times$\KW} (bnrelu3);

  \draw [->] (bnrelu3.north) -- (output);

  % braces
  \draw [decorate,decoration={calligraphic brace,amplitude=5pt,raise=18mm,mirror}, line width=1.25pt] (cam1.south)  -- node [midway,xshift=24mm] {\arrowFont $\times$3} (bnrelu2.north);

\end{tikzpicture}}
        \label{fig:model2}
      }
      &
      \subfigure[Diagram of an onset \texttt{\textbf{Stage}\textsubscript{$\shortdownarrow$}}. It is a modification of the stem, followed by convolutions that act like an MLP moving across time dimension \texttt{T'}. We add dropout to the MLP. The velocity \texttt{\textbf{Stage}\textsubscript{$V$}} is like \texttt{Stage\textsubscript{$\shortdownarrow$}}, except it has only $\{\times 1 \}$ CAM blocks and \texttt{17} input channels instead of \texttt{16}.]{
        \resizebox{0.45\linewidth}{!}{\newcommand{\horizMarginStage}{0mm}

\begin{tikzpicture}[node distance=3cm,
  every node/.style={fill=white, font=\sffamily},
  align=center]

  \node (input)[txt, minimum height=2mm]{Stage input\\[1mm]{\arrowFont B$\times$16$\times$\KW}};
  \node (conv1x1)[param, above of=input, yshift=-19mm]{Conv2d(1$\times$1)};
  \node (bnrelu1)[nonparam, above of=conv1x1, yshift=-14mm]{BN$\rightarrow$LReLU};

  \node (cam1)[param, above of=bnrelu1, yshift=-18mm]{CAM\textsubscript{1,2,3}(1$\times$11)};
  \node (bnrelu2)[nonparam, above of=cam1, yshift=-15mm]{BN$\rightarrow$LReLU};

  \node (convCollapse)[param, above of=bnrelu2, yshift=-18mm]{Conv2d(\textcolor{\violet}{\textbf{K}}$\times$3)};
  \node (bnrelu3)[nonparam, above of=convCollapse, yshift=-15mm]{BN$\rightarrow$LReLU$\rightarrow$Dropout};

  \node (convMLP1)[param, above of=bnrelu3, yshift=-20mm]{Conv2d(1x1)};
  \node (bnreludrop1)[nonparam, above of=convMLP1, yshift=-15mm]{BN$\rightarrow$LReLU$\rightarrow$Dropout};
  \node (convMLP2)[param, above of=bnreludrop1, yshift=-20mm]{Conv2d(1x1)};

  \node (sbn)[nonparam, above of=convMLP2, yshift=-14mm]{SBN\textsubscript{\textcolor{\violet}{\textbf{K}}}};
  \node (sigmoid)[nonparam, above of=sbn, yshift=-20mm]{Sigmoid};
  \node (output)[txt, above of=sigmoid, yshift=-20mm]{$\hat{\mathcal{R}}_{\shortdownarrow}$};

  \node (bottomMargin)[txt, below of=input, yshift=25mm, xshift=10mm]{};
  \node (leftMargin)[txt, left of=input, xshift=-\horizMarginStage]{};
  \node (rightMargin)[txt, right of=input, xshift=\horizMarginStage]{};

  % main arrow path
  \draw [->] (input.north) -- (conv1x1);
  \draw [->] (conv1x1.north) -- node[yshift=-1mm]{\arrowFont B$\times$12$\times$\KW} (bnrelu1);
  \draw [->,dashed,draw=black!50] (bnrelu1.north) -- (cam1);

  \draw [->] (cam1.north) -- node[yshift=-1mm]{\arrowFont B$\times$12$\times$\KW} (bnrelu2);
  \draw [->,dashed,draw=black!50] (bnrelu2.north) -- (convCollapse);

  \draw [->] (convCollapse.north) -- node[yshift=-1mm]{\arrowFont B$\times$200$\times$\oneW} (bnrelu3);

  \draw [->] (bnrelu3.north) -- (convMLP1);
  \draw [->] (convMLP1.north) -- node[yshift=-1mm]{\arrowFont B$\times$200$\times$\oneW} (bnreludrop1);
  \draw [->] (bnreludrop1.north) -- (convMLP2);
  \draw [->] (convMLP2.north) -- node[yshift=-1mm,xshift=8.5mm]{\arrowFont B$\times$\textcolor{\violet}{\textbf{K}}$\times$\oneW~$\equiv$~B$\times$\textcolor{\violet}{\textbf{K}}$\times$T'} (sbn);
  \draw [->] (sbn.north) -- (sigmoid);
  \draw [->] (sigmoid.north) -- (output);

  % braces
  \draw [decorate,decoration={calligraphic brace,amplitude=5pt,raise=18mm,mirror}, line width=1.25pt] (cam1.south)  -- node [midway,xshift=24mm] {\arrowFont $\times$3} (bnrelu2.north);

\end{tikzpicture}}  % textheight
        \label{fig:model3}
      }
    \end{tabular}
    \\
    \subfigure[
      Diagram of our proposed 3-stage \acl{OV} full architecture and training loss. Each \texttt{Stage\textsubscript{$\shortdownarrow$}$^{(i)}$} produces an onset piano roll $\hat{\mathcal{R}}_{\shortdownarrow}^{(i)}$, and successive stages refine the output via the residual connection ($+$ preserves shape). The velocity \texttt{Stage\textsubscript{$V$}} uses knowledge about onset locations from the final onset stage, and everything is trained jointly. See \cref{fig:qualitative}) for qualitative examples of $X$, $\hat{\mathcal{R}}_{\shortdownarrow}^{(3)}$ and $\hat{\mathcal{R}}_{V}$. \cref{subsec:training} presents the loss $l_{\shortdownarrow V}$.]{
      \resizebox{1\linewidth}{!}{\begin{tikzpicture}[node distance=3cm,
  every node/.style={fill=white, font=\sffamily},
  align=center]
  \node (input)[txt, minimum height=2mm]{$\!(X, \dot{X})$\\[1mm]{\arrowFont B$\times$2$\times$\FW}};
  \node (stem)[param, right of=input, xshift=-12mm] {Stem};

  \node (stage1)[param, right of=stem, xshift=-14mm, yshift=0mm] {Stage$_{\shortdownarrow}^{(1)}$};
  \node (stage2)[param, below of=stage1, xshift=0mm, yshift=20mm] {Stage$_{\shortdownarrow}^{(2)}$};
  \node (stage3)[param, below of=stage2, xshift=0mm, yshift=20mm] {Stage$_{\shortdownarrow}^{(3)}$};

  \node (plus1)[basecircle, right of=stage2, xshift=-14mm, yshift=0mm] {$+$};
  \node at (plus1 |- stage3) (plus2)[basecircle] {$+$};

  \node (cat)[basecircle, below of=plus2, xshift=0mm, yshift=20mm] {$\oplus$};

  \node (stageVel)[param, right of=cat, xshift=5mm, yshift=0mm] {Stage$_V$};

  \node (lossVel)[baseround, right of=stageVel, xshift=-1mm, yshift=0mm] {$l_{V'}(\mathcal{R}_{V_3}, \cdot)$};

  \node (loss3)[baseround, right of=plus2, xshift=-1.5mm, yshift=0mm] {$l_{BCE}(\mathds{1}_{\shortdownarrow_3}, \cdot)$};
  \node at (loss3 |- stage2)  (loss2)[baseround, xshift=0mm, yshift=0mm] {$l_{BCE}(\mathds{1}_{\shortdownarrow_3}, \cdot)$};
  \node at (loss3 |- stage1)  (loss1)[baseround, xshift=0mm, yshift=0mm] {$l_{BCE}(\mathds{1}_{\shortdownarrow_3}, \cdot)$};

  \node (onsetSum)[basecircle, right of=loss2, xshift=-4mm, yshift=0mm] {$+$};
  \node (lossFinal)[basecircle, right of=onsetSum, xshift=-2mm, yshift=-8mm] {$l_{\shortdownarrow V}$};

  \node (bottomMargin)[txt, below of=input, yshift=-5mm]{};

  %%%%%%%%%%%%%%%%%%%%%%%%%%%%%%%%%%%%%%%%%%%%
  \draw [->] (input.east) --  (stem);
  \draw [->] (stem.east) --  (stage1);
  \draw [->] (stem.south) |-  (stage2);
  \draw [->] (stem.south) |-  (stage3);
  \draw [->] (stem.south) |-  (cat);
  \draw [->] (stage1.east) -|  (plus1);
  \draw [->] (stage2.east) --  (plus1);
  \draw [->] (plus1.south) --  (plus2);
  \draw [->] (stage3.east) --  (plus2);
  \draw [->] (plus2.south) --  (cat);
  \draw [->] (cat.east) -- node [above=0.5mm, xshift=-1mm] {\arrowFont B$\times$17$\times$\KW} (stageVel);
  \draw [->] (stage1.east) -- node[above=0.5mm, xshift=5.1mm]{\arrowFont $\hat{\mathcal{R}}_{\shortdownarrow}^{(1)}$}  (loss1);
  \draw [->] (plus1.east) -- node[above=0.5mm]{\arrowFont $\hat{\mathcal{R}}_{\shortdownarrow}^{(2)}$}  (loss2);
  \draw [->] (plus2.east) --  node[above=0.5mm]{\arrowFont $\hat{\mathcal{R}}_{\shortdownarrow}^{(3)}$} (loss3);
  \draw [->] (stageVel.east) -- node[above=0mm, xshift=-1mm]{\arrowFont $\hat{\mathcal{R}}_V$}  (lossVel);
  \draw [->] (loss1.east) --  (onsetSum);
  \draw [->] (loss2.east) --  (onsetSum);
  \draw [->] (loss3.east) --  (onsetSum);

  \draw [->] (lossVel.east) --  (lossFinal);
  \draw [->] (onsetSum.east) --  (lossFinal);

\end{tikzpicture}}
      \label{fig:model4}
    }
  \end{tabular}
  \caption{\vspace{-0.55mm}Our proposed \ac{CNN}. Rank-4 tensor dimensions are \texttt{Batch$\times$Channel$\times$Height$\times$Width}. Design principles, interfaces and loss functions are described in \cref{sec:methodology}.}
  \label{fig:model}
  \vspace{-10mm}
\end{figure}

\vspace{3pt}
The complete \ac{CNN} is presented in \cref{fig:model}. We highlight the following design principles:
\vspace{3pt}

\paragraph{No recurrent layers} Motivated by \cite{regnet_nas,convnext}, we follow the established \ac{CNN} design of convolutional stem and body, followed by a fully connected head, making use of residual bottlenecks\cite{resnet}.
\paragraph{No pooling} Motivated by \cite{allconv}, all residual bottlenecks maintain activation shape, and conversion from input to output shape is done in a single depthwise convolution layer\cite{depthwise}, shown to be efficient and effective \cite{mobilenet}. Note that the convolutions in the input domain (spectrogram) have vertical dimensions but the convolutions in the output domain (piano roll) do not, since we assume that neighbouring frequencies are related but neighbouring piano keys aren't.
\paragraph{Multi-stage} Inspired by OpenPose\cite{openpose}, \ac{OV} features a series of residual stages that sequentially refine and produce the output. This is useful for real-time applications, since stages can be easily removed without need for retraining.
\paragraph{Temporal context} At its core, \ac{OV} features the \acfi{CAM} \cite{cam}, which is a residual bottleneck that combines time-dilated convolutions and channel attention\cite{squeeze}. Inspired by TCNs\cite{tcn} and Inception \cite{inception}, we aim to capture the temporal vicinity of an onset efficiently.
\paragraph{Model regularizers} At the input and before each output, \ac{OV} features \acfi{SBN}, i.e. one individual BN per vertical dimension \cite{batchnorm,sbn}. We add dropout\cite{dropout,bn_dropout_order} after the parameter-heavy layers. We use leaky ReLUs\cite{relu,leaky_relu} as nonlinearities.
\paragraph{Time locality} The time-derivative $\dot{X}$ is a handcrafted input feature that directly represents intensity variations.

\vspace{3pt}
During inference, \ac{OV} produces one piano roll per onset stage $(\hat{\mathcal{R}}_{\shortdownarrow}^{(1)}, \hat{\mathcal{R}}_{\shortdownarrow}^{(2)}, \hat{\mathcal{R}}_{\shortdownarrow}^{(3)})$ and one velocity piano roll $\hat{\mathcal{R}}_{V}$ (see \cref{fig:model4}). Then, our proposed decoder $\hat{\mathcal{S}} := dec_{\sigma, \rho, \mu}(\hat{\mathcal{R}}_{\shortdownarrow}^{(3)}, \hat{\mathcal{R}}_{ V})$ follows a simple heuristic: temporal Gaussian smoothing ({\it smooth}) with variance $\sigma^2$ followed by non-maximum suppression ({\it nms}), thresholding $\rho$ and shifting $\mu$, yielding the predicted score $\hat{\mathcal{S}}$ with note onsets and velocities:
\begin{align}
  \begin{split}
    \label{eq:decoder} 
    \hat{\shortdownarrow} &:= \big\{(k, t') :~ nms \big(smooth_{\sigma}(\hat{\mathcal{R}}_{\shortdownarrow}^{(3)}) \big){\scriptstyle (k, t')} \geq \rho \big\}\\
    \hat{\mathcal{S}} &:=\{(k_n,~   \hat{\mathcal{R}}_{V}(k_n, \! t_n'),~   \Delta_t t_n' \! + \! \mu) :~ (k_n, t_n') \in \hat{\shortdownarrow}_{\rho} \}
    \end{split}
\end{align}
The {\it nms} operation consists in zeroing out any $(k, t)$ entry that is strictly smaller than $(k, t + 1)$ or $(k, t - 1)$. The note events are read at the resulting locations, and shifted by a global constant $\mu$. In this work, we use the values $\sigma\!=\!1, \mu\!=\!-0.01s, \rho\!=\!0.74$, obtained via cross-validation of the trained \ac{CNN} on a subset of the MAESTRO validation split (note that this is different from the test split used for evaluation). While the optimal $\rho$ fluctuates during training, we found $\sigma\!=\!1,~\mu\!=\!-0.01s$ to be stable.

\begin{table*}[ht]
  \caption{Comparison of top-performing models in terms of  specifications (number of parameters for onset+velocity only, architecture and functionality) and performance (precision, recall, F\textsubscript{1}-score and MAESTRO version).\\[-3mm]}
  \label{tab:results}
  %% \vskip 0.15in
  \vspace{-2mm}
  %% \begin{center}
    \begin{small}
      \begin{sc}
        \resizebox{\linewidth}{!}{ \begin{tabular}{lcccccccccc}
            \toprule

            \multirow{2}[1]{*}{\textbf{Model}} &
            \multirowcell{2}{\bfseries Onset+Velocity \\ \bfseries \# Params} &
            \multirow{2}[1]{*}{\bfseries Architecture} &
            \multirow{2}[1]{*}{\bfseries Offset/Pedal?} &
            % & & &
            \multicolumn{3}{c}{\bfseries Onset (\%)} &
            %% & & &
            \multicolumn{3}{c}{\textbf{Onset+Velocity (\%)}}  &
            \multirowcell{2}{\bfseries MAESTRO \\ \bfseries version} \\

            \cmidrule(lr){5-7} \cmidrule(lr){8-10}
            & & & & {P} & {R} & {F\textsubscript{1}} & {P} & {R} & {F\textsubscript{1}} & \\
            \midrule
            \ac{OF} \cite{onsets_frames}               & 10M            & bi-RNN      & \cmark~/~\xmark & 98.27 & 92.61 & 95.32           & -     & -     & -      & v1 \\
            Regression \cite{kong21}                   & 12M            & bi-RNN      & \cmark~/~\cmark & 98.17 & 95.35 & 96.72           & -     & -     & -      & v2 \\
            Transformer \cite{hawthorne_transformer}   & --             & Transformer & \cmark~/~\xmark & -     & -     & 96.13           & -     & -     & -      & v3 \\
            \ac{OV} (ours)                             & {\bf 3.13M}    & CNN         & \xmark~/~\xmark & 98.58 & 95.07 & \textbf{96.78}  & 96.25 & 92.86 & 94.50  & v3 \\

            \bottomrule
          \end{tabular}}
      \end{sc}
    \end{small}
    %% \end{center}
    \vskip -2mm
\end{table*}

%%%%%%%%%%%%%%%%%%%%%%%%%
\subsection{Training}\label{subsec:training}
%%%%%%%%%%%%%%%%%%%%%%%%%

We train our \ac{CNN} to predict onset probability and velocity jointly via minimization of the following multi-task loss:
\begin{align*}
  \label{eq:ov_loss}
  %% \begin{align}
    l_{\shortdownarrow V} &(\mathds{1}_{\shortdownarrow_3}, (\hat{\mathcal{R}}_{\shortdownarrow}^{(1)},  \hat{\mathcal{R}}_{\shortdownarrow}^{(2)},  \hat{\mathcal{R}}_{\shortdownarrow}^{(3)}),  \! \mathcal{R}_{V_3}, \! \hat{\mathcal{R}}_{V}) :=\\[1mm]
    &\sum_{i}^3 l_{BCE}(\mathds{1}_{\shortdownarrow_3}, \hat{\mathcal{R}}_{\shortdownarrow}^{(i)}) + \lambda  \cdot   l_{V'}(\mathcal{R}_{V_3}, \! \hat{\mathcal{R}}_{V}) \text{, where}\\[1mm]
     l_{V'}&(\mathcal{R}_{V_3}, \! \hat{\mathcal{R}}_{V}) :=\\[1mm]
    &\big\langle \mathds{1}_{\shortdownarrow_3},~ \big(\mathcal{R}_{V_3} \! \cdot -log( \hat{\mathcal{R}}_{V}) \big) \! \big((1 \! - \! \mathcal{R}_{V_3}) \cdot  -log(1 \! - \! \hat{\mathcal{R}}_V)\big) \big\rangle
  %% \end{align}
\end{align*}
The $\mathds{1}_{\shortdownarrow_3}$ and $\mathcal{R}_{V_3}$ rolls are a straightforward modification of $\mathds{1}_{\shortdownarrow}$ and $\mathcal{R}_{V}$, where each active frame at $(k, t)$ is also extended into $t+1$ and $t+2$ (i.e. note onsets span 3 frames instead of one). This simple extension was crucial to achieve target performance, and combined with our decoder, allowed to bypass the need for elaborate decoding schemes as the ones discussed in \cite{kong21}. The masked loss $l_{V'}$ is a cross-entropy variant of the previously mentioned $l_V$, introduced in \cite{onsets_frames} and \cite{kong21}, that encourages to predict the right velocity only in the vicinity of onsets.

\vspace{3pt}
All model weights are initialized with the Gaussian-He distribution \cite{he_init}, and biases with 0, except the \ac{CAM} channel atention biases (right before the sigmoid), initialized with 1 to promote signal flow. We use the Adam optimizer with a decoupled weight decay \cite{adam,adamw} of 3$\times$10\textsuperscript{-4}, trained with random batches of 5-second segments (batch size 40, $\sim$14k batches per epoch) for $\sim$70K batches. For the learning rate, we start with a ramp-up from 0 to 0.008 across 500 batches, followed by cosine annealing with warm restarts \cite{sgdr}, using cycles of 1000 batches and decaying by 97.5\% after each cycle. BN/SBN momentum is 95\%, dropout 15\% and leaky ReLUs have a slope of 0.1. In $l_V'$, we use $\lambda\!=\!10$. To compensate that $\mathds{1}_{3\shortdownarrow}$ is sparse, we give positive entries a weight 8 times bigger than negative entries inside of $l_{BCE}(\mathds{1}_{3\shortdownarrow}, \cdot)$. Training speed was 1800 batches per hour on a 2080Ti NVIDIA GPU.

%%%%%%%%%%%%%%%%%%%%%%%%%
\subsection{Evaluation}
%%%%%%%%%%%%%%%%%%%%%%%%%

Following the same evaluation procedure as \ac{OF} \cite{onsets_frames}, \textsc{Regression} \cite{kong21} and \textsc{Transformer} \cite{hawthorne_transformer}, and applying standard metrics from \cite{ppt_eval} implemented in the \href{https://craffel.github.io/mir_eval/}{\texttt{mir\_eval}} library \cite{mir_eval}, we report precision (P), recall (R) and F\textsubscript{1}-score for the predicted {\it onsets}, considered correct if they are within 50ms of the ground truth. The {\it onset+velocity} evaluation, following \ac{OF}\cite[3.1]{onsets_frames}, has an added constraint: the predicted velocity must also be within 0.1 of the ground truth normalized between 0 and 1.

\vspace{3pt}
Note that the MAESTRO dataset is being actively extended and curated, presenting 3 versions so far. We report the respective versions in \cref{tab:results}, noting that versions 2 and 3 are almost identical, although comparisons across versions should be taken approximately.

%%%%%%%%%%%%%%%%%%%%%%%%%%%%%%%%%%%%%%%%%%%%%%%%%%%%%%%%%%%%%%%%%%%%%%%%%%%%%%%%%%%%%%%%%%%%%%%%%%%%%%
%%%%%%%%%%%%%%%%%%%%%%%%%%%%%%%%%%%%%%%%%%%%%%%%%%%%%%%%%%%%%%%%%%%%%%%%%%%%%%%%%%%%%%%%%%%%%%%%%%%%%%
\section{Experiments and Discussion}\label{sec:experiments}

We trained \ac{OV} on the MAESTRO v3 training split without any extensions or augmentations, achieving state-of-the-art performance in onset detection (see \cref{tab:results}). In the following we discuss some implications:

\vspace{3pt}
\paragraph{Temporal resolution} Our results seem to counter the need for increased temporal resolution expressed in \cite{kong21} (which use 8ms), showing that 24ms piano rolls coupled with our decoder presented in \cref{eq:decoder} are competitive.
%% \paragraph{Fast convergence} Training was performed on a 2080Ti NVIDIA GPU at 1800 batches (steps) per hour, reaching F\textsubscript{1}=96\% (onset) at $\sim$8k steps and target performance at $\sim$40k steps. This suggests {\it superconvergence}, possibly associated with the cyclical learning rate schedule as originally presented in \cite{superconvergence}.
\paragraph{Reduced memory footprint} \textbf{\cref{tab:results} reports model parameters for the components that are responsible exclusively for onset and frame detection} (\textsc{Transformer} has $\sim$54M parameters in total, but it transcribes everything jointly so it cannot be fairly compared). \ac{OV} outperforms the best alternative, \textsc{Regression}, with $\sim$4 times less parameters.

\paragraph{Affordable real-time inference} Bi-recurrent layers like the ones used in \ac{OF} and \textsc{Regression} are unsuited for real-time processing. \textsc{Transformer} took $\sim$380s to transcribe a 120s file on an Intel Xeon CPU (1 core), and $\sim$20s on a Tesla-T4 GPU (including offsets) when run on the official Colab implementation\footnote{\url{https://github.com/magenta/mt3}}. \ac{OV} took less than 2s to process the same file on an 8-core Intel i7-11800H CPU. Even accounting for the number of cores, \ac{OV} is approximately one order of magnitude faster than \textsc{Transformer}.

\paragraph{Conceptual simplicity} In essence, \ac{OV} revives the simplicity from \cite{cnn_maps16} by applying a feedforward \ac{CNN} to a discriminative task via piano rolls, followed by a simple decoding heuristic. Its architecture allows to remove onset stages without retraining, providing a flexible trade-off between runtime and performance with little added complexity.

\paragraph{Latency} The receptive field for our \ac{OV} proposed components is: \texttt{Stem}: 60 frames (1.44s), \texttt{Stage\textsubscript{$\shortdownarrow$}$^{(i)}$}: 99 frames ($\sim$2.38s), and \texttt{Stage\textsubscript{V}}: 35 frames ($0.84$s). This would theoretically impose a latency of over 9s, which is far from a responsive system. We informally note that the latency can be truncated without drastically affecting results (we used a latency of 4s in a live workshop), and encourage practical applications. We also note that our focus was on finding a \ac{CNN} with affordable inference and competitive performance, and we did not optimize for low receptive field, which may be obtainable with minor variations to the architecture (e.g. reducing the number of consecutive stages or \acp{CAM}).

%%%%%%%%%%%%%%%%%%%%%%%%%%%%%%%%%%%%%%%%%%%%%%%%%%%%%%%%%%%%%%%%%%%%%%%%%%%%%%%%%%%%%%%%%%%%%%%%%%%%%%
%%%%%%%%%%%%%%%%%%%%%%%%%%%%%%%%%%%%%%%%%%%%%%%%%%%%%%%%%%%%%%%%%%%%%%%%%%%%%%%%%%%%%%%%%%%%%%%%%%%%%%
\section{Conclusion and Future Work}\label{sec:conclusion}

We presented \acl{OV}, a convolutional setup that achieves real-time capabilities on modest commodity hardware without compromising performance, and with a substantially reduced size and temporal resolution. \ac{OV} achieves state-of-the-art performance in \ac{PPT} note {\it onset} detection, and establishes a good baseline on {\it onset+velocity} detection. Future work could include reducing the receptive field, extensions to {\it offset} and {\it pedal} detection, training and evaluation on different instruments, and analysis of design choices via ablation studies (e.g. number of stages and temporal resolution).

%%%%%%%%%%%%%%%%%%%%%%%%%%%%%%%%%%%%%%%%%%%%%%%%%%%%%%%%%%%%%%%%%%%%%%%%%%%%%%%%%%%%%%%%%%%%%%%%%%%%%%
%%%%%%%%%%%%%%%%%%%%%%%%%%%%%%%%%%%%%%%%%%%%%%%%%%%%%%%%%%%%%%%%%%%%%%%%%%%%%%%%%%%%%%%%%%%%%%%%%%%%%%
\section*{Acknowledgments}

A.F. wants to thank Jesús Monge Álvarez and Christian J. Steinmetz for their valuable feedback, the {\it Institut d'Estudis Baleàrics} for supporting this work with research grant 389062 INV-23/2021, and the International Max Planck Research School for Intelligent Systems (IMPRS-IS) for further support.

%%%%%%%%%%%%%%%%%%%%%%%%%%%%%%%%%%%%%%%%%%%%%%%%%%%%%%%%%%%%%%%%%%%%%%%%%%%%%%%%%%%%%%%%%%%%%%%%%%%%%%
%%%%%%%%%%%%%%%%%%%%%%%%%%%%%%%%%%%%%%%%%%%%%%%%%%%%%%%%%%%%%%%%%%%%%%%%%%%%%%%%%%%%%%%%%%%%%%%%%%%%%%
%% \clearpage
%% \section*{References}
% BIBLIOGRAPHY
\bibliographystyle{IEEEtran}
\bibliography{IEEEabrv,assets/references}

\begin{thebibliography}{10}
\providecommand{\url}[1]{#1}
\csname url@samestyle\endcsname
\providecommand{\newblock}{\relax}
\providecommand{\bibinfo}[2]{#2}
\providecommand{\BIBentrySTDinterwordspacing}{\spaceskip=0pt\relax}
\providecommand{\BIBentryALTinterwordstretchfactor}{4}
\providecommand{\BIBentryALTinterwordspacing}{\spaceskip=\fontdimen2\font plus
\BIBentryALTinterwordstretchfactor\fontdimen3\font minus
  \fontdimen4\font\relax}
\providecommand{\BIBforeignlanguage}[2]{{%
\expandafter\ifx\csname l@#1\endcsname\relax
\typeout{** WARNING: IEEEtran.bst: No hyphenation pattern has been}%
\typeout{** loaded for the language `#1'. Using the pattern for}%
\typeout{** the default language instead.}%
\else
\language=\csname l@#1\endcsname
\fi
#2}}
\providecommand{\BIBdecl}{\relax}
\BIBdecl

\bibitem{ismir2009}
J.~Downie, D.~Byrd, and T.~Crawford, ``Ten years of {ISMIR}: Reflections on
  challenges and opportunities.'' \emph{ISMIR Proc.}, pp. 13--18, 01 2009.

\bibitem{amt13}
E.~Benetos, S.~Dixon, D.~Giannoulis, H.~Kirchhoff, and A.~Klapuri, ``Automatic
  music transcription: Challenges and future directions,'' \emph{Journal of
  Intelligent Information Systems}, vol.~41, 12 2013.

\bibitem{driedger}
J.~Driedger, ``Processing music signals using audio decomposition techniques,''
  doctoral thesis, FAU Erlangen-N{\"u}rnberg, 2016.

\bibitem{dl_nature}
Y.~LeCun, Y.~Bengio, and G.~Hinton, ``Deep learning,'' \emph{Nature}, vol. 521,
  pp. 436--44, 05 2015.

\bibitem{maps}
V.~Emiya, N.~Bertin, B.~David, and R.~Badeau, ``{MAPS} - a piano database for
  multipitch estimation and automatic transcription of music (research
  report),'' 2010.

\bibitem{a_maps}
A.~Ycart and E.~Benetos, ``{A-MAPS}: Augmented {MAPS} dataset with rhythm and
  key annotations,'' \emph{19th ISMIR Conference}, 09 2018.

\bibitem{smd}
M.~M{\"u}ller, V.~Konz, W.~Bogler, and V.~Arifi-M{\"u}ller, ``Saarland music
  data ({SMD}),'' \emph{Late-Breaking and Demo Session of the 12th {ISMIR}},
  2011.

\bibitem{musicnet}
J.~Thickstun, Z.~Harchaoui, and S.~M. Kakade, ``Learning features of music from
  scratch,'' in \emph{ICLR}, 2017.

\bibitem{maestro}
C.~Hawthorne, A.~Stasyuk, A.~Roberts, I.~Simon, C.-Z.~A. Huang, S.~Dieleman,
  E.~Elsen, J.~Engel, and D.~Eck, ``Enabling factorized piano music modeling
  and generation with the {MAESTRO} dataset,'' in \emph{International
  Conference on Learning Representations}, 2019.

\bibitem{giantmidi}
Q.~Kong, B.~Li, J.~Chen, and Y.~Wang, ``{GiantMIDI-Piano}: A large-scale {MIDI}
  dataset for classical piano music.'' \emph{ISMIR Trans.}, pp. 87--98, 2022.

\bibitem{kong21}
Q.~Kong, B.~Li, X.~Song, Y.~Wan, and Y.~Wang, ``High-resolution piano
  transcription with pedals by regressing onset and offset times,''
  \emph{IEEE/ACM Trans. Audio, Speech and Lang. Proc.}, vol.~29, p.
  3707–3717, oct 2021.

\bibitem{cnn_maps16}
R.~Kelz, M.~Dorfer, F.~Korzeniowski, S.~B{\"{o}}ck, A.~Arzt, and G.~Widmer,
  ``On the potential of simple framewise approaches to piano transcription,''
  in \emph{ISMIR Proceedings}, august 2016, pp. 475--481.

\bibitem{leaf}
N.~Zeghidour, O.~Teboul, F.~de~Chaumont~Quitry, and M.~Tagliasacchi, ``{LEAF}:
  A learnable frontend for audio classification,'' in \emph{International
  Conference on Learning Representations}, 2021.

\bibitem{bracewell}
R.~Bracewell, \emph{The Fourier Transform and its Applications}, 2nd~ed.\hskip
  1em plus 0.5em minus 0.4em\relax Tokyo: McGraw-Hill Kogakusha, Ltd., 1978.

\bibitem{panns}
Q.~Kong, Y.~Cao, T.~Iqbal, Y.~Wang, W.~Wang, and M.~D. Plumbley, ``{PANN}s:
  Large-scale pretrained audio neural networks for audio pattern recognition,''
  \emph{IEEE/ACM Transactions on Audio, Speech, and Language Processing},
  vol.~28, pp. 2880--2894, 2020.

\bibitem{onsets_frames}
C.~Hawthorne, E.~Elsen, J.~Song, A.~Roberts, I.~Simon, C.~Raffel, J.~H. Engel,
  S.~Oore, and D.~Eck, ``{O}nsets and {F}rames: Dual-objective piano
  transcription.'' in \emph{ISMIR}, 2018, pp. 50--57.

\bibitem{adv_of}
J.~Kim and J.~Bello, ``Adversarial learning for improved onsets and frames
  music transcription,'' in \emph{ISMIR Proceedings}, 2019, pp. 670--677.

\bibitem{of_analysis}
K.~Cheuk, Y.~Luo, E.~Benetos, and D.~Herremans, ``Revisiting the {O}nsets and
  {F}rames model with additive attention,'' in \emph{Proceedings of the
  International Joint Conference on Neural Networks (IJCNN)}, 2021.

\bibitem{lstm}
S.~Hochreiter and J.~Schmidhuber, ``Long short-term memory,'' \emph{Neural
  computation}, vol.~9, pp. 1735--80, 12 1997.

\bibitem{birnn}
M.~Schuster and K.~Paliwal, ``Bidirectional recurrent neural networks,''
  \emph{IEEE Trans. on Sig. Proc.}, vol.~45, no.~11, pp. 2673--2681, 1997.

\bibitem{hawthorne_transformer}
C.~Hawthorne, I.~Simon, R.~Swavely, E.~Manilow, and J.~H. Engel,
  ``Sequence-to-sequence piano transcription with transformers,'' in
  \emph{ISMIR Proceedings}, november 2021, pp. 246--253.

\bibitem{transformers}
A.~Vaswani, N.~Shazeer, N.~Parmar, J.~Uszkoreit, L.~Jones, A.~N. Gomez, L.~u.
  Kaiser, and I.~Polosukhin, ``Attention is all you need,'' in \emph{Advances
  in Neural Information Processing Systems}, vol.~30, 2017.

\bibitem{kelz21}
R.~Kelz and G.~Widmer, ``Nonlinear denoising, linear demixing,'' in \emph{I
  (Still) Can't Believe It's Not Better! NeurIPS 2021 Workshop}, 2021.

\bibitem{dds_overfitting}
L.~S. Mart\'{a}k, R.~Kelz, and G.~Widmer, ``Balancing bias and performance in
  polyphonic piano transcription systems,'' \emph{Frontiers in Signal
  Processing}, vol.~2, 2022.

\bibitem{adsr_kelz}
R.~Kelz, S.~B{\"o}ck, and G.~Widmer, ``Deep polyphonic {ADSR} piano note
  transcription,'' \emph{ICASSP}, pp. 246--250, 2019.

\bibitem{invertible}
R.~Kelz and G.~Widmer, ``Towards interpretable polyphonic transcription with
  invertible neural networks,'' in \emph{ISMIR Proceedings}, 2019.

\bibitem{no_subtask}
K.~Cheuk, Y.~Luo, E.~Benetos, and D.~Herremans, ``The effect of spectrogram
  reconstruction on automatic music transcription: An alternative approach to
  improve transcription accuracy,'' in \emph{2020 25th International Conference
  on Pattern Recognition (ICPR)}, 2021, pp. 9091--9098.

\bibitem{pytorch}
A.~Paszke, S.~Gross, F.~Massa, A.~Lerer, J.~Bradbury, G.~Chanan, T.~Killeen,
  Z.~Lin, N.~Gimelshein, L.~Antiga, A.~Desmaison, A.~Kopf, E.~Yang, Z.~DeVito,
  M.~Raison, A.~Tejani, S.~Chilamkurthy, B.~Steiner, L.~Fang, J.~Bai, and
  S.~Chintala, ``{PyTorch}: An imperative style, high-performance deep learning
  library,'' in \emph{Advances in Neural Information Processing Systems}, 2019,
  pp. 8024--8035.

\bibitem{mel}
S.~S. Stevens and J.~E. Volkmann, ``The relation of pitch to frequency: A
  revised scale,'' \emph{American Journal of Psychology}, vol.~53, p. 329,
  1940.

\bibitem{regnet_nas}
I.~Radosavovic, R.~P. Kosaraju, R.~Girshick, K.~He, and P.~Dollar, ``Designing
  network design spaces,'' in \emph{{CVPR} Proceedings}, June 2020.

\bibitem{convnext}
Z.~Liu, H.~Mao, C.-Y. Wu, C.~Feichtenhofer, T.~Darrell, and S.~Xie, ``A convnet
  for the 2020s,'' \emph{CVPR PRoceedings}, 2022.

\bibitem{resnet}
K.~He, X.~Zhang, S.~Ren, and J.~Sun, ``Deep residual learning for image
  recognition,'' in \emph{CVPR Proceedings}, Jun. 2016, pp. 770--778.

\bibitem{allconv}
J.~T. Springenberg, A.~Dosovitskiy, T.~Brox, and M.~A. Riedmiller, ``Striving
  for simplicity: The all convolutional net,'' in \emph{ICLR}, 2015.

\bibitem{depthwise}
F.~Chollet, ``{Xception}: Deep learning with depthwise separable
  convolutions,'' in \emph{CVPR Proceedings}, 2017, pp. 1800--1807.

\bibitem{mobilenet}
A.~G. Howard, M.~Zhu, B.~Chen, D.~Kalenichenko, W.~Wang, T.~Weyand,
  M.~Andreetto, and H.~Adam, ``{MobileNets}: Efficient convolutional neural
  networks for mobile vision applications,'' \emph{CoRR}, 2017.

\bibitem{openpose}
Z.~{Cao}, G.~{Hidalgo Martinez}, T.~{Simon}, S.~{Wei}, and Y.~A. {Sheikh},
  ``{OpenPose}: Realtime multi-person {2D} pose estimation using part affinity
  fields,'' \emph{IEEE Trans. on Pattern Analysis and Machine Intelligence},
  2019.

\bibitem{cam}
J.~Zhang, Z.~Chen, and D.~Tao, ``Human keypoint detection by progressive
  context refinement,'' in \emph{ICCV Workshop: COCO Keypoint Detection
  Challenge Track}, 10 2019.

\bibitem{squeeze}
J.~Hu, L.~Shen, and G.~Sun, ``Squeeze-and-excitation networks,'' in \emph{CVPR
  Proceedings}, 2018, pp. 7132--7141.

\bibitem{tcn}
S.~Bai, J.~Z. Kolter, and V.~Koltun, ``An empirical evaluation of generic
  convolutional and recurrent networks for sequence modeling.'' \emph{CoRR},
  2018.

\bibitem{inception}
C.~Szegedy, W.~Liu, Y.~Jia, P.~Sermanet, S.~Reed, D.~Anguelov, D.~Erhan,
  V.~Vanhoucke, and A.~Rabinovich, ``Going deeper with convolutions,'' in
  \emph{CVPR Proceedings}, 2015, pp. 1--9.

\bibitem{batchnorm}
S.~Ioffe and C.~Szegedy, ``Batch normalization: Accelerating deep network
  training by reducing internal covariate shift,'' in \emph{ICML Proceedings},
  2015, p. 448–456.

\bibitem{sbn}
S.~Chang, H.~Park, J.~Cho, H.~Park, S.~Yun, and K.~Hwang, ``Subspectral
  normalization for neural audio data processing,'' \emph{ICASSP}, 2021.

\bibitem{dropout}
N.~Srivastava, G.~Hinton, A.~Krizhevsky, I.~Sutskever, and R.~Salakhutdinov,
  ``Dropout: A simple way to prevent neural networks from overfitting,''
  \emph{Journal of Machine Learning Research}, vol.~15, no.~56, pp. 1929--1958,
  2014.

\bibitem{bn_dropout_order}
X.~Li, S.~Chen, X.~Hu, and J.~Yang, ``Understanding the disharmony between
  dropout and batch normalization by variance shift,'' in \emph{CVPR
  Proceedings}, 2019, pp. 2682--2690.

\bibitem{relu}
V.~Nair and G.~E. Hinton, ``Rectified linear units improve restricted boltzmann
  machines,'' in \emph{ICML Proceedings}, 2010, p. 807–814.

\bibitem{leaky_relu}
A.~L. Maas, A.~Y. Hannun, and A.~Y. Ng, ``Rectifier nonlinearities improve
  neural network acoustic models,'' in \emph{ICML Proceedings}, 2013.

\bibitem{he_init}
K.~He, X.~Zhang, S.~Ren, and J.~Sun, ``Delving deep into rectifiers: Surpassing
  human-level performance on imagenet classification,'' in \emph{ICCV
  Proceedings}, 2015, pp. 1026--1034.

\bibitem{adam}
D.~P. Kingma and J.~Ba, ``Adam: A method for stochastic optimization,'' in
  \emph{ICLR Proceedings}, 2015.

\bibitem{adamw}
I.~Loshchilov and F.~Hutter, ``Decoupled weight decay regularization,'' in
  \emph{ICLR Proceedings}, 2019.

\bibitem{sgdr}
\vspace{0mm}I. Loshchilov and F.~Hutter, ``{SGDR}: Stochastic gradient descent
  with warm restarts,'' in \emph{ICLR Proceedings}, 2017.

\bibitem{ppt_eval}
M.~Bay, A.~F. Ehmann, and J.~S. Downie, ``Evaluation of multiple-{F0}
  estimation and tracking systems,'' in \emph{ISMIR Proc.}, 2009, pp. 315--320.

\bibitem{mir_eval}
C.~Raffel, B.~McFee, E.~J. Humphrey, J.~Salamon, O.~Nieto, D.~Liang, and
  D.~P.~W. Ellis, ``\texttt{mir\_eval}: A transparent implementation of common
  {MIR} metrics,'' in \emph{ISMIR Proceedings}, 2014, pp. 367--372.

\end{thebibliography}
\end{document}